\documentclass[11pt, titlepage]{article}
\usepackage{palatino}
\usepackage[affil-it]{authblk}
\usepackage{graphicx}
\usepackage[left=1in, right=1in, top=1in]{geometry}
\usepackage{hyperref}
\usepackage{natbib}
\usepackage{fancyhdr}
\usepackage{setspace} 
\usepackage{lastpage}

\usepackage{authblk}

\newcommand{\mms}{m\&m's\textsuperscript{\tiny{\textregistered}}~}
\newcommand{\mmsnospace}{m\&m's\textsuperscript{\tiny{\textregistered}}}

\title{Introducing Bayesian Analysis with \mmsnospace: \\ an active-learning exercise for undergraduates}

\author[a,b,c]{Gwendolyn Eadie
	\thanks {corresponding author: eadieg@uw.edu} }

\author[a,b,c]{Daniela Huppenkothen
	\thanks{dhuppenk@uw.edu} }
\author[d]{Aaron Springford
	\thanks{stats@aaronspringford.com} }

\author[b,e,f]{Tyler McCormick
	\thanks {tylermc@u.washington.edu} }

\affil[a]{Department of Astronomy, Box 351580, University of Washington, Seattle, WA, 98195-1580}
\affil[b]{eScience Institute, Campus Box 351570, U.W.,  3910 15th Ave NE, Seattle, WA 98195}
\affil[c]{DIRAC Institute, Box 351580,  University of Washington, Seattle, WA, USA 98195-1580}
\affil[d]{Weyerhaeuser Company, 220 Occidental Ave S, Seattle, WA 98104}
\affil[e]{Department of Statistics, Box 354322, University of Washington, Seattle, WA 98195-4322}
\affil[f]{Department of Sociology, University of Washington, Seattle, WA }

\date{Accepted to the Journal of Statistics Education on 28 March 2019}

\begin{document}

\maketitle

\begin{abstract}

We present an active-learning strategy for undergraduates that applies Bayesian analysis to candy-covered chocolate \mmsnospace. The exercise is best suited for small class sizes and tutorial settings, after students have been introduced to the concepts of Bayesian statistics. The exercise takes advantage of the non-uniform distribution of \mms colours, and the difference in distributions made at two different factories. In this paper, we provide the intended learning outcomes, lesson plan and step-by-step guide for instruction, and open-source teaching materials. We also suggest an extension to the exercise for the graduate-level, which incorporates hierarchical Bayesian analysis.

\vspace{2ex}

\sc{keywords: \small{\emph{Education, Bayesian Methods, Inference, Active-learning, Eliciting Priors}}}

\end{abstract}

\section{Introduction}

We have developed an active-learning exercise for upper-year undergraduates that applies Bayesian analysis to \mms candy. It is a fun activity that can be completed in a tutorial or small classroom setting within a 50-80 min class.  Part of the exercise relies on the fact that \mms are made at two different factories, and that the colour distributions produced at these factories are different. The exercise may also be extended to the graduate-level as a way to introduce and practically apply hierarchical Bayesian analysis (Section~\ref{sec:advanced} and Appendix~\ref{appendix}).

In short, the activity involves giving each student a bag of \mms (their data), and guiding them through an exercise to perform Bayesian analysis in order to:

\begin{enumerate}
	\item \emph{Infer the probability of drawing a blue \mms from a new bag of \mmsnospace, given a likelihood, prior, and their data.}
	\item \emph{Predict the factory from which the \mms were produced, based on the posterior distribution for the entire class' data}.
\end{enumerate}

The exercise is primarily meant for undergraduates who have been introduced to Bayesian statistics, but who have not yet applied Bayesian inference to a real problem. The exercise is also relevant to anyone new to Bayesian inference who has a quantitative background. The mathematics in the exercise involves probability density functions that are both analytic and readily available in many open-source software packages. Thus,  it is straightforward for students to simplify the likelihood times prior on the fly, and also write computer code to plot their results immediately in class. In the Github repository associated with this paper, we also provide \textbf{R} and \textbf{Python} scripts for the instructor and/or student to complete the exercise. If class time allows, then it might be beneficial for students to write their own scripts. 

In the next section of this paper, Section~\ref{sec:motivation}, we describe the pedagogical motivation and development of the Bayesian \mms exercise. In Section~\ref{sec:colours}, we describe the true colour distribution of \mms as produced by the MARS company, and how this relates to the exercise. The lesson plan is presented in Section~\ref{sec:lessonplan}, which includes an overview (\ref{sec:overview}), a list of intended-learning outcomes (\ref{sec:ILOs}), and a detailed step-by-step outline (\ref{sec:outline}) with suggested discussion questions and strategies for implementation of the exercise in classrooms. Next, we present the actual posterior distributions found by an undergraduate class and by a seminar for graduate students, postdocs, and faculty. Following this, we briefly describe and provide links to the publicly available software instructors might like to use while running our active-learning exercise (Section~\ref{sec:opensource}). Finally, we present two extensions to the \mms exercise that are intended for more advanced (e.g., graduate-level) students (Section~\ref{sec:advanced} and Appendix~\ref{appendix}).

\section{Motivation and Development}\label{sec:motivation}	

By the time students are introduced to Bayesian analysis, they have usually had several classes on the frequentist perspective. Therefore,  it is important to provide students with a concrete way to conceptualize the Bayesian framework, and this can be achieved through active-learning. The \mms exercise presented in this paper provides such an activity.

Author GE, in collaboration with AS, originally created the Bayesian \mms lesson and activity for a guest lecture in a third-year undergraduate astronomy class focused on statistics. The students had already been exposed to the Bayesian perspective but needed an interactive example to solidify the basic concepts. The class enrollment was approximately 10 students, and classtime was 80 minutes. Thus, the learning environment was well-suited for an active-learning exercise that would engage students and allow them to apply what they had learned about Bayesian statistics to a tangible example. 

While searching for an interactive learning tool for introductory Bayesian inference, GE found blog posts describing how the colour distribution of \mms can be used as a teaching tool for frequentist statistics\footnote{\tiny\url{https://blogs.sas.com/content/iml/2017/02/20/proportion-of-colors-mandms.html}}\footnote{\tiny\url{https://joshmadison.com/2007/12/02/mms-color-distribution-analysis/}}\footnote{\tiny\url{https://katedegner.wordpress.com/tag/mm-color-distribution/}}. There is an \mms exercise using Bayes' theorem in \cite{downey2013}, but the content is quite brief and the intended-learning outcome (ILO) is somewhat unclear. The goal for the latter is to predict whether an \mms came from a bag produced in 1994 or 1996 (the colour distributions changed in 1995, when blue \mms were introduced). Thus, it seems the ILO for the latter exercise was not to introduce Bayesian inference, but possibly to practice calculating probabilities. 

A handful of publications describing learning activities with \mms also exist \citep[e.g.,][]{alexander1994,  alexsherri1994, SherriJoyce1994, fricker1996,  1998Dyck, 2006LinSanders, 2013FroelichRobert, 2013schwartz}). As a group, these articles cover a variety of topics in statistics, such as regression and correlation, analysis of variance, sampling distribution of the mean, design of experiments,  and chi-squared goodness-of-fit tests. Other activities have also been suggested for more general mathematics education, from population modeling \citep[e.g.,][]{2009Winkel} to memoryless processes and hypergeometric functions \citep[e.g.,][]{2017Badinski}. 

\cite{2009AlbertsRossman} --- a workbook intended for an entire introductory course on statistics through the Bayesian perspective ---  includes a Bayesian exercise using \mmsnospace, entitled, ``What proportion of \mms are brown?''. This exercise resides in the Chapter ``Learning about models using Bayes' rule'', and is not meant to be a comprehensive, stand-alone activity about Bayesian inference. The activity first states that a bag of \mms has 3 brown candies out of 10 (this is the ``real'' data). Next, a table is presented that shows mock \mms data that were simulated from four different factory models. The students are tasked with comparing the ``real'' \mms data to the simulations, along with their prior assumptions, in order to determine the best factory model given their data. Thus, the exercise is not a thorough introduction to Bayesian inference, and instead plays a small part in a workbook that has `` [...] its emphasis on active learning and its use of the Bayesian viewpoint to introduce the basic notions of statistical inference'' \citep[Preface, ][]{2009AlbertsRossman}.

In summary, a comprehensive Bayesian example using \mms candies does not seem to exist in the education literature nor is one publicly available online, even though \mms have been used as a teaching tool for statistics since the 1990s. Thus, we designed this comprehensive active-learning exercise with \mms to help students learn and practice the concepts of Bayesian analysis. 
 
GE first implemented our Bayesian \mms activity in a third-year undergraduate class and received very positive feedback from students. The exercise has also been presented and informally relayed at conferences in both astronomy and statistics, with consistently enthusiastic responses. We have since formalized the lesson plan, learning outcomes, and teaching materials, and made them available through this manuscript. We have also developed and included an extension to the \mms exercise for more advanced classes that can be used to introduce concepts in hierarchical Bayesian analysis  (Section~\ref{sec:advanced} and Appendix~\ref{appendix}).

Our goal is that the \mms exercise may be widely used and improved upon by the greater statistics community and other quantitative disciplines that teach and use Bayesian analysis.

\section{The intriguing colour distributions of \mms}\label{sec:colours}

\begin{figure}
	\centering
	\includegraphics[scale=0.25]{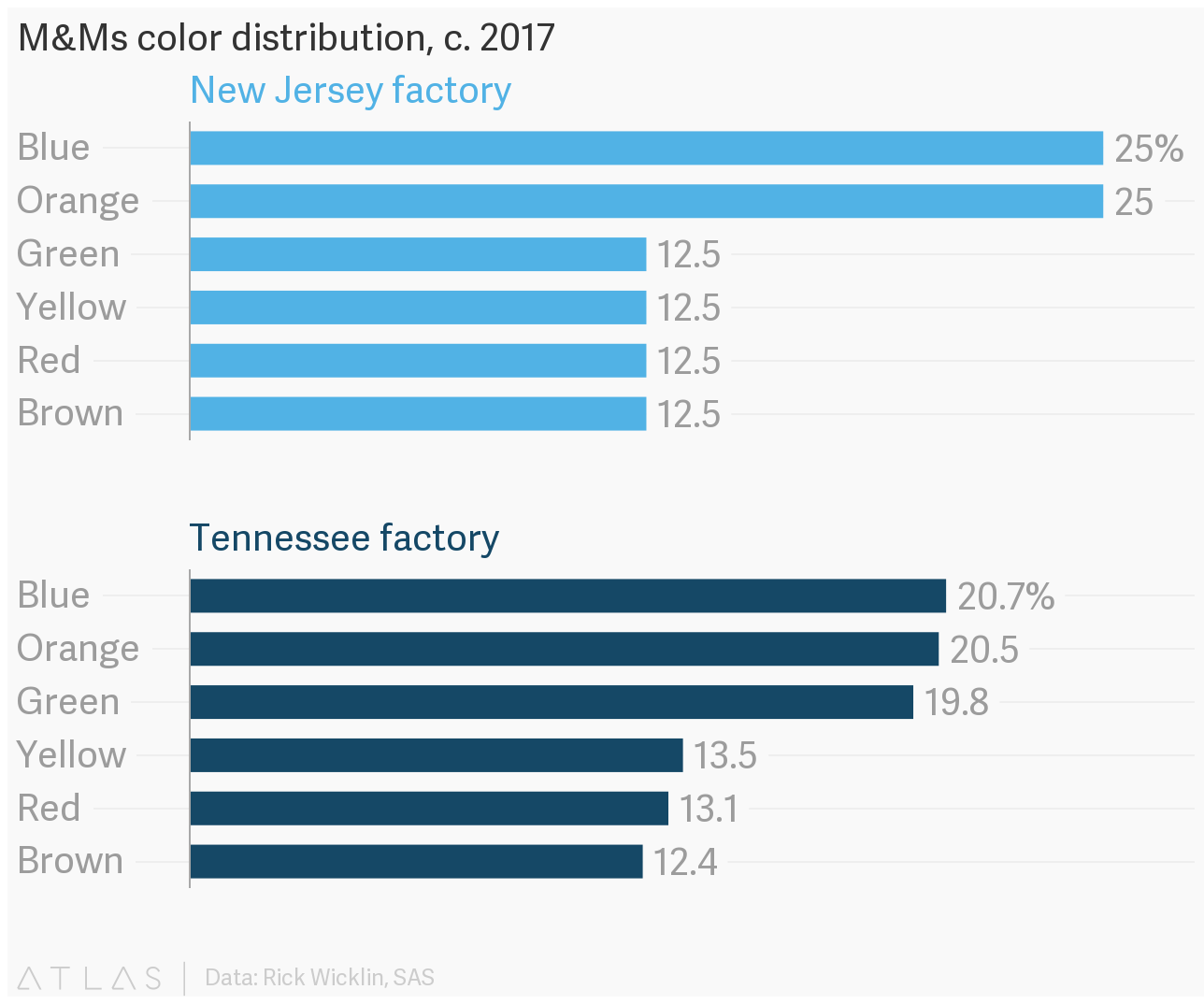}
	\caption{The \mms colour distributions produced at the New Jersey and Tenessee factories, respectively. The percentage of blue, orange, and green \mms differ the most between the two factories. This chart originally appears in \cite{quartz2017}, and was created by \cite{atlaschart}, using data from \cite{wicklin2017}.}\label{fig:twofactories}
\end{figure}

The active-learning exercise we present here takes advantage of an important fact about the production of \mmsnospace: two factories of the MARS Company (one in  Hackettstown, New Jersey, the other in Cleveland, Tennesee) make \mmsnospace, but these factories produce different distributions of \mms colours! As shown in Figure~\ref{fig:twofactories}, the New Jersey and Tennessee factories make significantly different percentages of blue, orange, and green \mmsnospace. The New Jersey and Tennessee factories make 25\% and 20.7\% blue \mms respectively, and this is the colour we use throughout the exercise. Of course, orange or green could instead be used and should give similar results.

Due to the design of our activity, it is imperative to keep the true colour distribution a secret from students until the end of the exercise. The exercise assumes most students have prior knowledge about \mms (e.g., they have eaten them), but that they are unaware of the nonuniform colour distribution. In our experience, students are surprised to find out at that the \mms colour distributions are actually nonuniform, and that they also differ between the two factories. We have conducted this exercise with New Jersey and Tennessee \mmsnospace, and the posterior distributions predicted the correct factory in both cases. We used peanut \mmsnospace, but the exercise should work with traditional \mms too (and would also be more allergy-friendly). Admittedly, the exercise also relies on the factories not changing their production lines for the foreseeable future.

\section{Lesson Plan}\label{sec:lessonplan}

The entire lesson consists of interactive lecturing, discussion, and active-learning. We recommend not handing out the \mms candy too early in the lesson, lest students become distracted, or worse, eat the evidence before it is recorded! In Sections~\ref{sec:overview}--\ref{sec:outline}, we provide an overview for the exercise, ILOs, and detailed steps and discussion questions.
\vspace{1ex}

\subsection{Overview}\label{sec:overview}

The lesson begins with a ``hook'' to engage students. A bag of \mms is shown to the class, and they are asked to think about the distribution of colors inside. Next, they are asked how they would predict the percentage of blue \mms produced at the factory, using only a single bag of \mmsnospace. The instructor leads a discussion and points out the benefits of using Bayesian inference, because it can incorporate 
the students' prior knowledge about \mmsnospace. Finally, the students are presented with the exercise: Use Bayes' theorem and a single bag of \mms as data to predict the percentage of blue \mms produced at the factory.

To help the students start the problem, Bayes theorem is reviewed and the instructor helps students decide what likelihood will be appropriate for their study. The instructor also elicits prior information from the class, and helps students quantify the prior distribution based on this information. (Of course, this could instead be presented in reverse order, with first eliciting prior information and then introducing a likelihood.) The instructor can either select a method of prior elicitation in advance, or can present different options to the class for consideration and discussion if wanted.

Next, students work out the product of the likelihood and prior to obtain an analytic form of the posterior distribution. If there is time and computer resources available, they may also write and prepare a short computer script to plot this distribution. Alternatively, they may use the Jupyter Notebook provided as supplementary material to this manuscript. Next, each student is given a bag of \mms to open and inspect. They record the total number of \mmsnospace and the number of each colour, and then eat the evidence if they so wish.

Students now plot the posterior distribution given their data using either their computer script or the one provided, and then discuss the results with a partner. Following this, the instructor pools the \mms data from the entire class and creates a posterior distribution for all of the \mmsnospace. A discussion and question period is then led about the effect of more data on the posterior distribution.

Next, the surprise twist is revealed (Figure~\ref{fig:twofactories}), and students are asked to infer which factory produced their bag of \mmsnospace, using their posterior distribution. They can also compare their inference to what they would infer given the entire class' \mmsnospace. After the students have arrived at an answer, the instructor provides the factory codes for the two factories. Students can check the lot number on the back of their \mms packages for verification. In our experience, a class size of ten or more students seems to provide enough data for the mode of the posterior to accurately predict the percentage of blue \mms and the factory from which the candy originated. 

It should be noted that our exercise may also provide an opportunity to compare the Bayesian formalism to the frequentist approach. Exactly how to make these connections will depend on the students' level and depth of exposure to both topics, so we leave this to the discretion of the instructor. However, we strongly suggest that any such comparisons be done in a follow-up lesson or in a homework assignment, as adding this material to our exercise would be too much to fit into one class.

\subsection{Intended Learning Outcomes}\label{sec:ILOs}

This lesson assumes that students have been exposed to the idea of probability distributions and have seen Bayes' theorem, at least in passing. They should also have an understanding of parameters and their role in defining the shape of a probability distribution. As mentioned previously, we also assume that students do not have prior knowledge about the surprise twist mentioned in Section~\ref{sec:colours}.

The intended learning outcomes (ILOs) are as follows. By the end of this lesson, students should be able to:
\begin{enumerate}

	 \item Recall Bayes' theorem and identify the likelihood, prior, and posterior.
	 
	 \item Recognize and quantify prior information.
	 
	 \item State the conjugate prior to the binomial distribution, and list the hyperparameters.
	  
	 \item Understand the term hyperparameters.
	 
	 \item Calculate the posterior distribution for the probability of drawing a blue \mms from a new candy bag, given the \mms data and the prior distribution.

	\item Write a working computer script to plot and compare the prior distribution to the posterior distribution obtained with their data (if computing resources are available).

	 \item Perform inference based on the Bayesian posterior distribution.
	 
	 \item Know the appropriate terminology for reporting results from Bayesian analysis.
	 
	 \item Connect this simple example to similar real-world practical applications from the students' domain of study. 
	 % The last item above is related to a conversation between Gwen and Daniela: as the last part of the exercise, students should be encouraged to think about a real-world application from their domain where they could apply their newly learned knowledge. This is of course only the case if students aren't statistics students. We might want to suggest an example like this somewhere in the paper?
	
\end{enumerate}

\subsection{Detailed Steps and Discussion Questions}\label{sec:outline}

 In many of the steps listed below, we include suggested discussion questions as bullets.

\begin{enumerate}
	\item  As a ``hook'', show a bag of \mms to the class and explain that they are going to infer the probability of getting a blue \mms using Bayes' Theorem. Each person will get a bag of \mms as their data, and they will use this data with a probability model to infer the percentage of blue \mms produced in the factory.
	
	\item\label{item:questions} Ask students prompting questions to help them formulate a quantitative picture of the problem.
	
		\begin{itemize}
			\item What kind of data will we have when we open the \mms bags? Is it numerical? Categorical? Continuous?
			
			\item What's one way to estimate the percentage of blue \mms in the bag? Are there any pitfalls to the approach?
			
			\item If you observe zero blue \mms in your bag, then what is your estimate of the fraction of blue \mms in the population?  Is this realistic?
			
			\item How will we record the data from the \mms bags?
			
		\end{itemize}
	
		\item Review Bayes' Theorem and its constituent parts.

		\item Lead a discussion about what likelihood might best describe the probability of drawing a blue \mms from the bag, given their answers to the questions in Item \ref{item:questions} above. Ask for student input and guide the class to use a binomial distribution. 
		\begin{itemize}
			\item What assumptions are made by the likelihood?
			\item Are we sampling with replacement or without?
			\item After recording the data, can we eat it or should we wait until the analysis is complete?
		\end{itemize}
	
	\item Return to their answers in Item~\ref{item:questions} to talk about prior information and to develop an approximate distribution for their prior belief in the percentage of blue \mmsnospace \footnote{ A well-defined method of prior elicitation from the literature could be used to guide this part. A couple of entry points to the literature are \citet{garthwaite2005statistical} and \citet{kadane1998experiences}. Prior elicitation is a broad topic, but one worth introducing to the students if possible; in our experience, prior elicitation is rarely covered in coursework, but can be vitally important to statistical practice.}. Some questions for eliciting prior information could include:
		\begin{itemize}
			\item How do you think \mms are made in the factory?
			
			\item Do you think the manufacturing process affects the proportion of blue \mms in a typical bag?
			
			\item Do you think \mms are well mixed before they are put into bags?
			
			\item What do you think the percentage of blue \mms is, at the factory? Do you think every bag will have this exact percentage?
			
		\end{itemize}
	
	\item Help the students quantify their prior information. \label{quanitfyprior}
		\begin{itemize}
			\item Where do you think most of the prior probability should be? 
			
			\item What parts of parameter space would you like to apply a very low probaiblity?
			
			\item Sketch out a distribution that encompasses your belief about the percentage of blue \mms in the bag.
			
		\end{itemize}
		
	\item If conjugate priors are not a new concept to students, then briefly discuss why they are useful. If conjugate priors are new, introduce the beta distribution and its parameters, and explain its conjugacy to the binomial distribution.
						
	\item  Explore the properties of the beta distribution given its hyperparameters. If computer resources are available, students can interactively plot the prior distribution until they achieve a prior distribution that quantifies their prior belief about the proportion of blue \mmsnospace. 

		\begin{itemize}
			\item What happens to the beta distribution as both hyperparameters go to one? Do you recognize this distribution?
			
			\item What happens when the hyperparameters are equal?
			
			\item What values of the hyperparameters best approximate the prior distribution sketched in Item~\ref{quanitfyprior}?
			
		\end{itemize}
	\newpage
	\item Choose the hyperparameters, thus defining the prior distribution. 
			
		\begin{itemize}
			\item  Why is it important that we define the prior distribution before looking at the data (i.e., before opening the bag of \mmsnospace)? 
			
			\item How many \mms worth of information does your prior contain?
		\end{itemize}

	\item As a class, decide on the hyperparameter values for the instructor's prior distribution (which will be used with all the class' data). One way to achieve class agreement is for the instructor to interactively plot the prior distribution for hyperparameter values, suggested by the students, until a consensus is reached.\label{choosehyper}
		\begin{itemize}
			\item What is the mean of the prior distribution that was chosen?
		
			\item How does the prior distribution compare to a uniform distribution? How does the prior distribution compare to a simple assumption that the probability of drawing a blue \mms is $\frac{1}{6}$?
		\end{itemize}
	
	\item Students work out the product of likelihood and prior, and identify the kernel of this distribution. If computer resources are available, they write a short script to take in their data and plot the posterior distribution. Test their script with some toy data.
	
	\item Open the bags of \mms and look at the data. Make sure students record not only the number of blue \mmsnospace, $y$, but also the total number of all \mmsnospace, $n$.
		\begin{itemize}
			\item How do the number of blue \mms in your bag compare to your neighbor's bag?
			\item Do you think there are enough \mms in one bag to correctly infer the percentage of blue \mms produced at the factory?

		\end{itemize}
	
	\item  (If computer resources are not available, then skip to the next step). Students plot the posterior distribution given their data and prior information, and compare it to the prior distribution.
		\begin{itemize}
		\item Where is most of the probability for the percentage of blue \mmsnospace?
		
		\item How has the posterior changed from the prior distribution?
		
		\item Is this the result you expected, given that there are six different colours of \mmsnospace?
		
		\item How does your posterior distribution compare to your neighbor's?
		\end{itemize}
	
	\item Instructor compiles the class' data for $y$ and $n$. 

	\begin{itemize}
		\item What assumptions have we made when combining data from different bags?
		\item Do you have any predictions about how the posterior distribution might change in light of more data?

	\end{itemize}
	
	\item Instructor produces a plot comparing the class posterior distribution to the class' prior distribution.
			\begin{itemize}
		\item Where is most of the probability for the percentage of blue \mms?
		
		\item How does the posterior distribution for the data from the whole class compare to your own?
		
		\item How does the prior compare to the posterior, now that there is more data?
		
		\item Has the shape of the posterior changed with more data? Did it change in the way you expected?
		
		\item What is the expected value (mean) and variance of the new posterior distribution for the percentage of blue \mmsnospace?
		
		\item Is this the result you expected for the percentage of blue \mmsnospace, given that there are six different colours of \mmsnospace? What might you infer about the colour distribution of \mms from these results?
			
		\item Discuss the best way to report and display these results.
		
	\end{itemize}

	\item (Optional) Repeat the analysis for each of the other colours (red, orange, green, and yellow) for each colour of \mmsnospace. Repeating the analysis may help students master the practical components of the exercise which are generally useful, such as identifying the variable of interest, checking their computer code, and presenting results through graphics.

	\item Reveal the surprise twist: Show the colour distributions presented in Figure~\ref{fig:twofactories}, and ask students to discuss with one another and infer from the posterior distribution which factory their \mms came from.
		
	\item  Once they have performed inference, then they can look at the back of the \mms bag to see if their inference is correct. The lot number will show either CLV (Cleveland, Tennessee factory) or HKP (Hackettstown, New Jersey).

\end{enumerate}

\section{Mathematics for the \mms exercise}

Bayes' theorem states that the probability of $\theta$ given $y$ is
\begin{equation}
	p(\theta|y) = \frac{p(y|\theta)p(\theta)}{p(y)},
\end{equation}
where $p(y|\theta)$ is the likelihood. For the probability of drawing a blue \mms from the bag, we use a binomial distribution for the likelihood 
\begin{equation}\label{eq:binomial}
	p(y|\theta) \propto \theta^y (1-\theta)^{n-y},
\end{equation}
where $y$ is the number of successes (blue \mms), $n$ is the total number of \mms drawn from the bag, and $\theta$ is the percentage of blue \mms produced at the factory. 

More details about eliciting prior information from the students is presented in Section~\ref{sec:outline}. After eliciting prior information and sketching an approximate prior distribution, we set out to quantify this information. We use the conjugate prior to Equation~\ref{eq:binomial}, the beta distribution, for the prior on $\theta$: 
\begin{equation}
\label{eq:beta}
p(\theta) \propto \theta^{\alpha-1}(1-\theta)^{\beta - 1},
\end{equation}
with hyperparameters $\alpha$ and $\beta$. Equation~\ref{eq:beta} not only simplifies the \mms example for a time-constrained class, but also provides the opportunity to review the concept of a conjugate prior.

Now that the prior has been defined, the posterior distribution is proportional to
\begin{equation}
p(\theta|y) \propto \theta^{y + \alpha -1} (1 - \theta)^{n-y + \beta -1}.
\end{equation}
We find it useful for students to simplify the likelihood and prior on their own, and then use think-pair-share to identify the form of the posterior distribution. Once students have recognized that the posterior is also a beta distribution, they may use software to plot the posterior distribution given their data (i.e., their \mmsnospace).

\newgeometry{top=1in, left=1in, right=1in, bottom=0.9in}
\section{Implementation and Posterior Distributions}

\begin{figure}
	\centering
	\includegraphics[scale=0.5, trim=0cm 0cm 0cm 0cm]{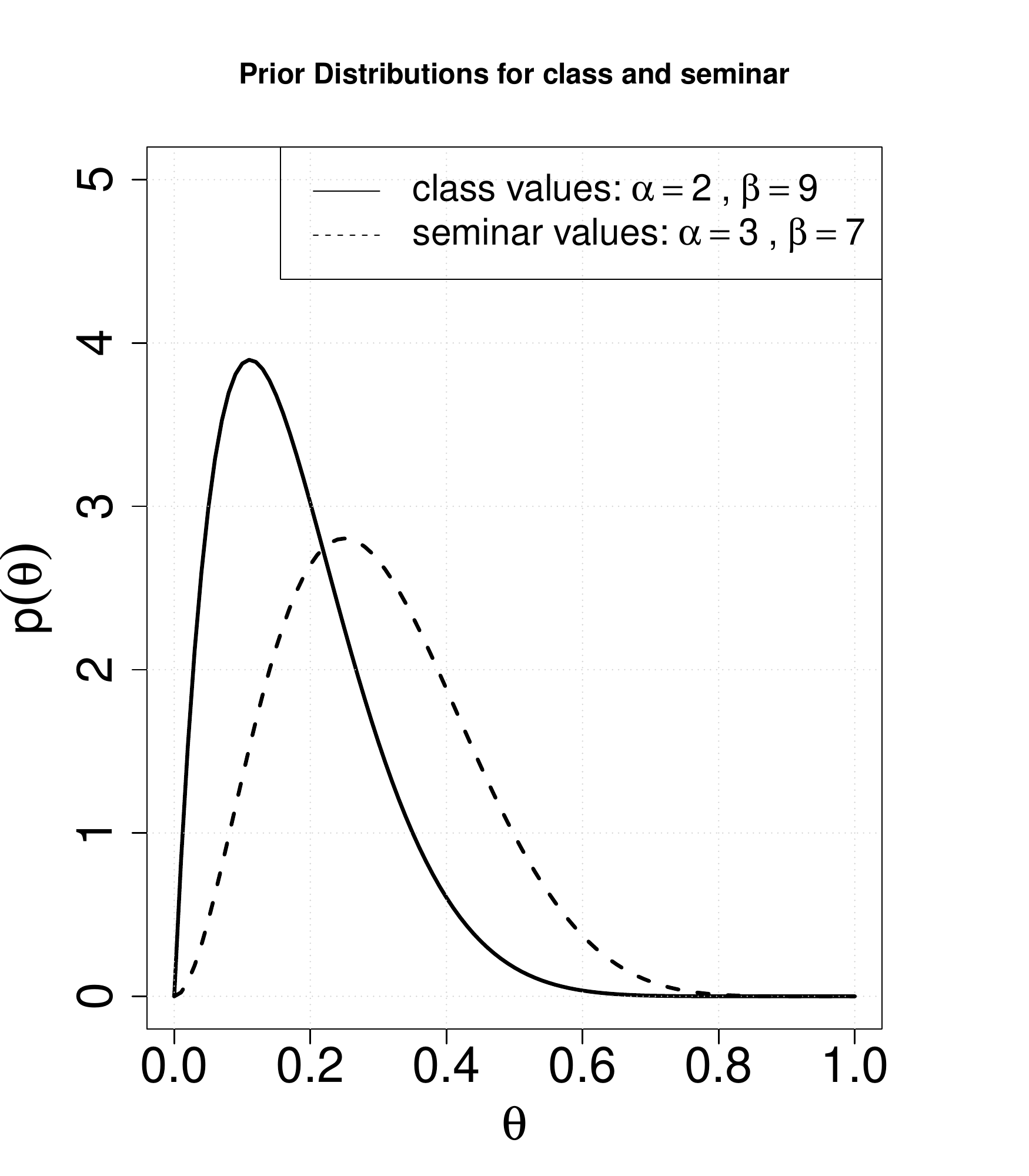}
	\caption{The prior distribution for the probability of drawing a blue \mms from a bag of \mmsnospace, as determined by the class of undergraduates (solid  curve) and by the seminar of graduate students, postdocs, and faculty (dashed curve). In general, students believed it was less likely to draw a blue \mms than did the seminar members.}\label{fig:prior}
\end{figure}

We implemented the exercise presented in this paper in a third-year undergraduate astronomy class, and in a seminar for graduate students, postdocs, and faculty at the Institute for Data Intensive Research in Astrophysics \& Cosmology (DIRAC). In this section, we show the posterior distribution that resulted from both instances.

In our class of approximately 10 students, it was agreed that a beta distribution with hyperparameters $\alpha=2$ and $\beta=9$ best described their prior information about the probability of drawing a blue \mms from a bag. We reached this agreement through open discussion and the questions suggested in Items~\ref{quanitfyprior}-\ref{choosehyper} of Section~\ref{sec:outline}. In the seminar, the graduate students, postdocs, and faculty, used  hyperparameters $\alpha =3$ and $\beta=7$. Both prior distributions are shown in Figure~\ref{fig:prior}.

\begin{figure}
	\centering
	\includegraphics[scale=0.15]{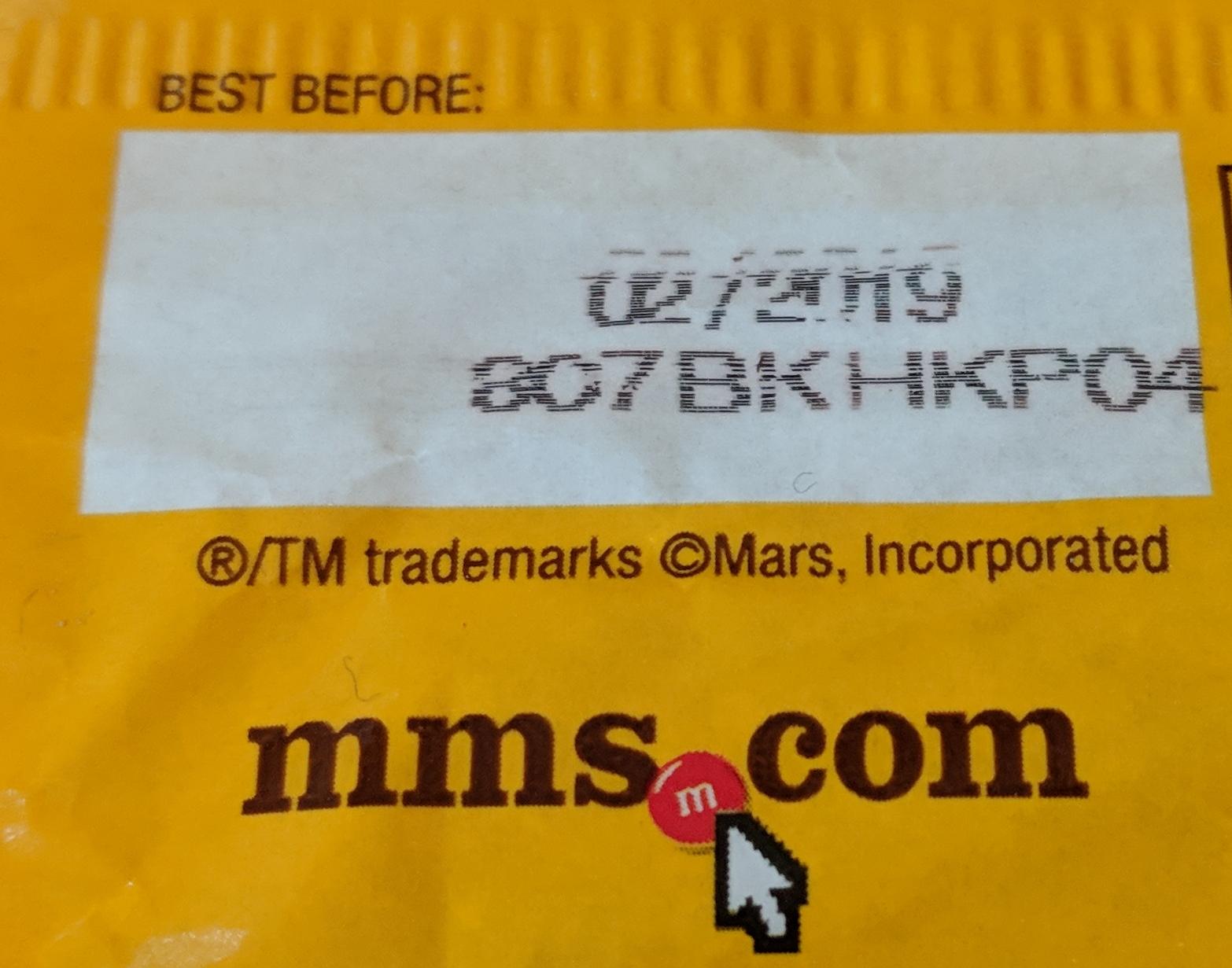}
	\caption{A picture of the lot number from a package in the third-year undergraduate class. The code HKP indicates the package came from the Hackettstown, New Jersey factory.}\label{fig:HKP}
\end{figure}

The posterior distribution for the first author's data (i.e., one bag of \mmsnospace), and the posterior distribution for the class' data from all bags, are shown with the prior distribution in Figure~\ref{fig:posterior}. The first author's data are included in the supplementary material. The ensemble data consisted of 100 \mmsnospace, 25 of which were blue. The class inferred from the total posterior distribution and its mean that the \mms must have originated from the New Jersey factory. We checked the back of the packages, and indeed they showed the code HKP, indicating they came from Hackettstown, NJ (Figure~\ref{fig:HKP}).

For the class, we had purchased individual packages of peanut \mms  from a local store in Seattle, WA. For the seminar, however, we expected more than twenty people--- so instead we bought a large box of peanut \mms packages from Costco\textsuperscript{\tiny{\textregistered}}. 
\restoregeometry

\begin{figure}
	\centering
	\includegraphics[scale=0.6, trim=2cm 0cm 0cm 0cm]{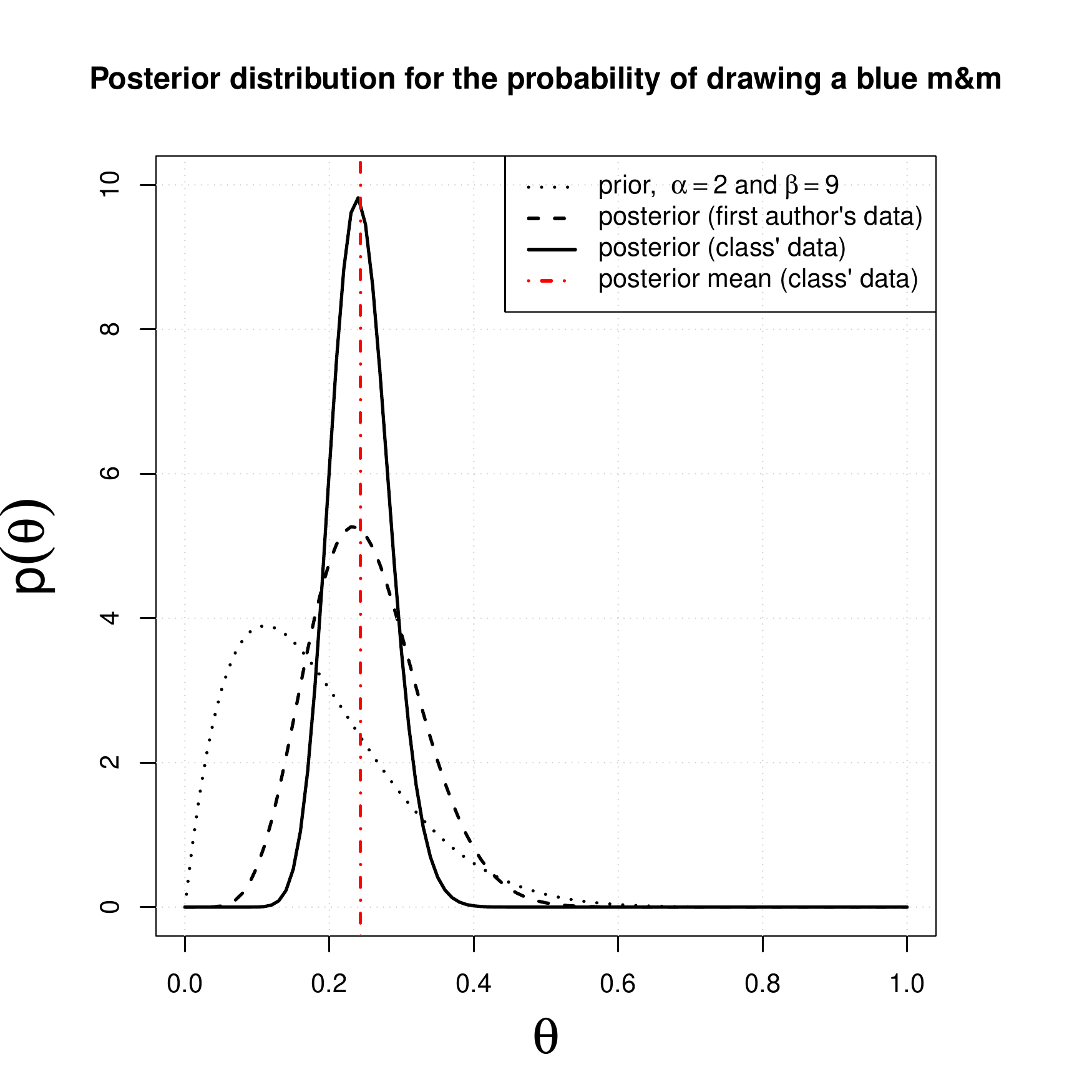}
	\caption{Posterior distributions for instructor's data (dashed curve), and for the combined data from the entire class (solid curve). The prior distribution (dotted curve) and posterior mean for the class' posterior distribution (red vertical dashed-dotted line) are also shown for reference. The \mms used were produced in the Hackettstown, New Jersey factory (HKP).}
	\label{fig:posterior}
\end{figure}

\begin{figure}[b]
	\centering
	\includegraphics[scale=0.7, trim=2cm 0cm 0cm 0cm]{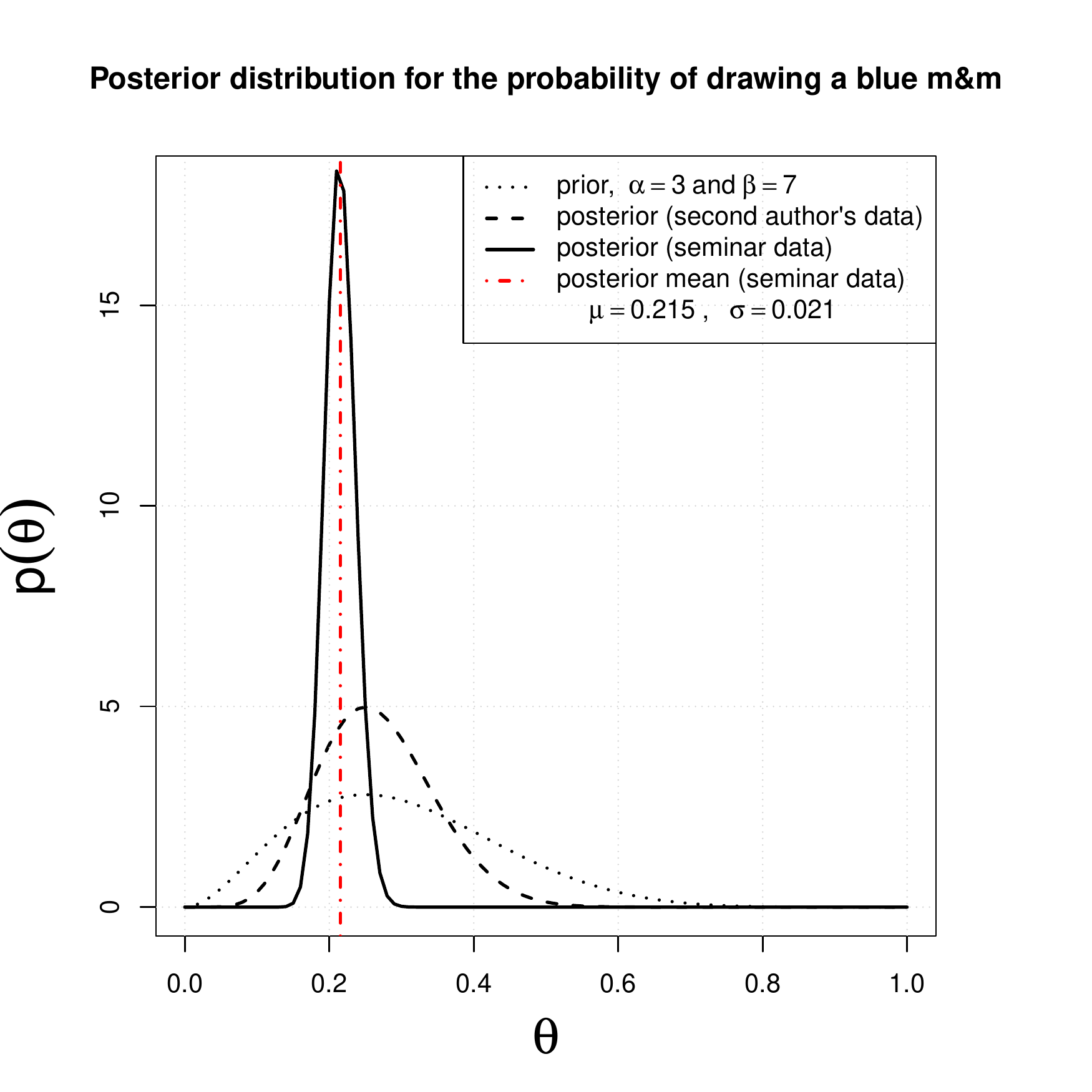}
	\caption{Posterior distributions for the DIRAC Institute seminar and the first author's bag of \mmsnospace. The \mms used for this seminar were produced in the Cleveland, Tennessee factory (code CLV), and the data used to generate this figure are included as supplemental material.}
	\label{fig:posteriorseminar}
\end{figure}
\begin{figure}[b]
	\centering
	\includegraphics[scale=0.15]{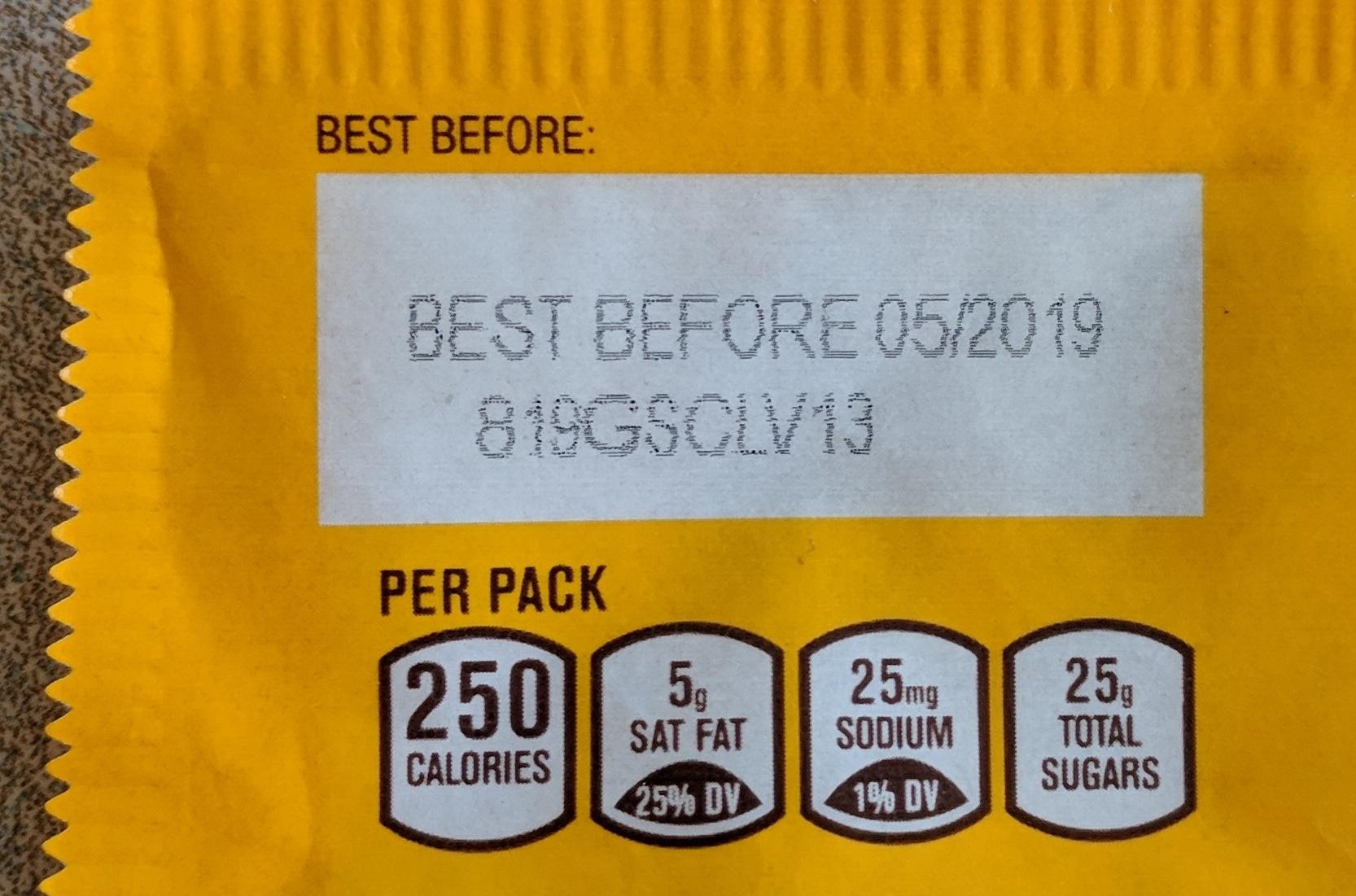}
	\caption{A picture of the lot number from a package from the seminar. The code CLV indicates the package came from the Cleveland, Tennessee factory.}\label{fig:CLV}
\end{figure}
\clearpage
The results from the DIRAC seminar are presented in Figure~\ref{fig:posteriorseminar}. In this case, the first author was very surprised to see a different result than the undergraduate class, as they had expected the packages to also come from New Jersey. As it turned out, the \mms packages in the large box were produced in Cleveland, Tennessee (Figure~\ref{fig:CLV})! The data are included in a supplementary csv file entitled \emph{DIRAC\_mmsTallyCLVTennessee}. Interestingly, Seattle, WA has packages from both factories, and perhaps other North American cities do too. This bodes well for instructors looking to purchase \mms from a specific factory (e.g., when they wish to implement the advanced version of this exercise  in Section~\ref{sec:advanced}).

\section{Open Source Scripts}\label{sec:opensource}
Ideally, the students should be given a chance to write computer scripts to plot the prior and posterior distributions in this exercise, in their chosen software (e.g., R, Python, SAS, etc.). If classtime is short, then it might expedite the exercise to use the pre-authored scripts we provide on the Github repository accompanying this paper\footnote{\url{https://github.com/gweneadie/BayesianMandMs}}.

\section{Extensions for Advanced Classes}\label{sec:advanced}

For higher-level (e.g., graduate-level) classes, there are several ways to extend this exercise to introduce and encompass more advanced concepts. We explored extensions that use the masses of the \mmsnospace, that incorporate quality control levels at the factory, that consider multiple colours, or that include the colour distributions of \mms produced at different factories. The latter two most naturally follow from the original exercise, and so we describe these two ideas in more detail below.

A straightforward extension generalizes the case of two possible outcomes of the data (blue versus not-blue \mmsnospace) to the case of six possible outcomes (i.e., blue, orange, green, yellow,  red, or brown). By including a parameter for the probability of each colour, a multi-dimensional posterior distribution is naturally introduced. More specifically, modeling all six colours of \mms simultaneously involves the multinomial distribution and its conjugate prior, the Dirichlet distribution --- both of which have a vast range of applications in domains with categorical data (e.g., topic modeling; see e.g.,~ \citealt{blei2012probabilistic} for an introduction). 

An even more advanced extension not only incorporates  multiple colours within a bag, but also the fact that \mms bags may come from one of two factories, and that the two factories (Tennessee and New Jersey) produce different colour distributions. Incorporating the individual bags, class collection of bags, and the two factories into the exercise provides a natural basis for introducing and applying Bayesian hierarchical modeling and mixture models.

Instead of creating a joint data set from all bags in the class and obtaining a posterior distribution (which assumes all bags of \mms originated in the same factory), the students are asked to write down a joint model for all bags, with latent variables to be inferred along with the overall distributions. Students must jointly infer the colour distributions produced at each factory, the latent assignment of each bag to a factory, and the mixture proportions of factories within the set of bags in the class. The hierarchical nature of this model is in the joint inference of the latent mixture assignments with the mixture proportions. In addition, the colour distributions are inferred as global population parameters, while the joint information from all bags will shrink the posteriors on the latent mixture assignment variables. The specifics of the hierarchical model for the \mms exercise are shown in Appendix~\ref{appendix}.

The hierarchical version of the \mms exercise is appealing because the variables' true values can be known by the instructor; the colour proportions from each factory are relatively well known (Figure~\ref{fig:twofactories}, as long as the MARS company does not change their practices), and the instructor can deliberately set the mixture proportions of the bags (depending on the availability of \mms from both factories). In the context of a longer course on Bayesian analysis, this exercise can also be used to introduce related advanced concepts such as probabilistic graphical models as well as Markov Chain Monte Carlo methods and Gibbs sampling.

\section{Conclusion \& Future Work}

Many successful active-learning activities that use \mms exist for topics in frequentist statistics and mathematics education. The literature, however, was lacking a Bayesian \mms example. We hope that this exercise fills the gap. 

Introducing Bayesian analysis to undergraduate students, regardless of their field of study, requires a discussion of Bayes' theorem, probability distributions, and prior distributions.  Examples that are meant to illuminate these concepts often rely on objects or systems such as playing cards, dice, or urns. Although these examples can be useful and familiar, we feel they are overused and have become repetitive both for students and instructors. Most undergraduates have already seen examples using playing cards, dice, and urns in high school-level mathematics classes, so it would be nice to have an entirely different type of example to entice student engagement. Students must also have the opportunity to apply what they are learning and, in Bayesian analysis in particular, develop practical skills. We hope that this \mms example provides an enticing (and tasty!) alternative to traditional examples. 

It would be beneficial and probably more cost effective if instructors could source \mms from the two MARS factories directly. We have not contacted the MARS Company about direct purchasing, but should they read this manuscript, we are happy to discuss setting up a program for \mms  discounted offers in the name of statistical education.

\section*{Acknowledgements}
The authors thank the editors and the anonymous referees whose comments and suggestions helped improve this manuscript.

This research was funded by an eScience Institute Postdoctoral Fellowship to the first author, G.~Eadie. G.~Eadie and T.~McCormick acknowledge support from the three Moore, Sloan, and Washington Research Foundations at the eScience Institute, University of Washington. G.~Eadie and D.~Huppenhothen acknowledge support from the University of Washington Institute for Data Intensive Research in Astrophysics and Cosmology (DIRAC). The DIRAC Institute is supported through generous gifts from the Charles and Lisa Simonyi Fund for Arts and Sciences, and the Washington Research Foundation. The authors would like to thank the DIRAC Institute members for their useful discussions relating to this exercise, and for their willingness to eat \mms on behalf of pedagogical development. The authors would also like to thank C.~Morrison for identifying the location of the lot code on the candy packages during the DIRAC seminar.

\bibliographystyle{apalike}
\bibliography{MandMrefs}

\newpage

\appendix
\section{Details of the Hierarchical Bayesian Model}\label{appendix}

\begin{figure}[h]
	\centering
	\includegraphics[scale=0.8, trim=2cm 0cm 0cm 0cm]{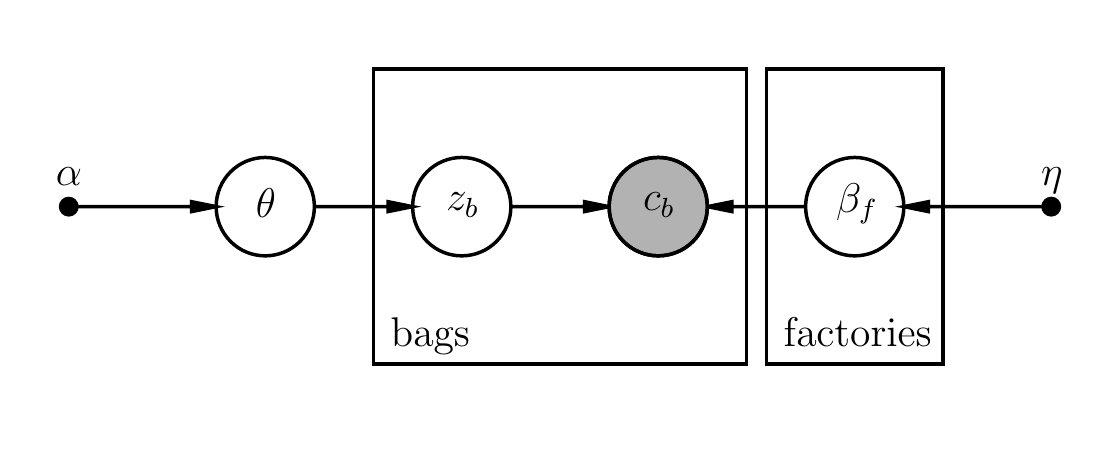}
	\caption{A probabilistic graphical model describing the hierarchical approach to modeling $b$ bags of \mms from different factories $f = 1,2, \dots $. The vector $c_b$ denotes the number of \mms of each of the six colours in a single bag $b$. $z_b$ are the latent assignments of each bag to a factory, $\theta$ indicates the mixture proportions for the two factories, and $\beta_f$ are the colour distributions for each factory. Finally, $\alpha$ and $\eta$ denote the hyperparameters for the Dirichlet priors on $\theta$ and $\beta_f$.}
	\label{fig:graphmodel}
\end{figure}

In the hierarchical version of the exercise, there are two sets of parameters $\beta_1$ and $\beta_2$ (one for each factory, $f$), and each set represents the fraction of \mms per colour that are produced in Factories 1 and 2. A Dirichlet prior is placed on each set of parameters, with hyperparameters $\eta$. In addition, each bag $b$ out of the total sample of $B$ bags is given a latent categorical variable $z_b$ that assigns the bag to Factory 1 or Factory 2. This variable is drawn from a categorical distribution describing the probability of drawing a bag from either factory. The parameters for the mixture proportions is denoted as $\theta$. The latter, too, has a Dirichlet prior, with hyperparameters $\alpha$. A graphical version of this model is shown in Figure \ref{fig:graphmodel}.

The full posterior distribution for the categorical mixture model described in Figure \ref{fig:graphmodel}  can be written as 

\begin{equation}
p(\{z_b\}_{b=1}^{B}, \{\beta_f\}_{f=1}^2, \theta | \{c_b\}_{b=1}^{B}, \eta, \alpha) = p(\theta | \alpha) \prod_{f=1}^{2}\left[ p(\beta_f | \eta) \right] \prod_{b=1}^B\left[ p(c_b | z_b, \beta_f) p(z_b | \theta) \right] / Z,
\end{equation}

\noindent where

\begin{eqnarray}
\theta \sim \mathrm{Dirichlet}(\alpha), \\ \nonumber
\beta_f \sim \mathrm{Dirichlet}(\eta), \\ \nonumber
z_b \sim \mathrm{Categorical}(\theta),~ and \\ \nonumber
c_b \sim \mathrm{Multinomial}(z_b, \beta_f). \nonumber \;
\end{eqnarray}

\vspace{10ex}

\end{document}